\documentclass[10pt]{article}

\usepackage{amsmath}
\usepackage{amssymb}

\usepackage{cite}

\usepackage{hyperref}

\usepackage{lineno}
\usepackage{graphicx}

\usepackage{microtype}
\DisableLigatures[f]{encoding = *, family = * }

\usepackage[labelfont=bf,labelsep=period,justification=raggedright]{caption}

\makeatletter
\renewcommand{\@biblabel}[1]{\quad#1.}
\makeatother

\date{}

\begin{document}

\begin{flushleft}
{\Large
\textbf{Using synchronous Boolean networks to model several phenomena of collective behavior}
}
\\
Stepan Kochemazov$^{1}$,
Alexander Semenov$^{1,\ast}$ 
\\
\bf{1} ISDCT SB RAS, Irkutsk, Russia
\\
$\ast$ E-mail: biclop.rambler@yandex.ru
\end{flushleft}

\section*{Abstract}
In this paper, we propose an approach for modeling and analysis of a number of phenomena of collective behavior. By collectives we mean multi-agent systems that transition from one state to another at discrete moments of time. The behavior of a member of a collective (agent) is called conforming if the opinion of this agent at current time moment conforms to the opinion of some other agents at the previous time moment. We presume that at each moment of time every agent makes a decision by choosing from the set $\{0,1\}$ (where $1$-decision corresponds to action and $0$-decision corresponds to inaction). In our approach we model collective behavior with synchronous Boolean networks. We presume that in a network there can be agents that act at every moment of time. Such agents are called instigators. Also there can be agents that never act. Such agents are called loyalists. Agents that are neither instigators nor loyalists are called simple agents. We study two combinatorial problems. The first problem is to find a disposition of instigators that in several time moments transforms a network from a state where a majority of simple agents are inactive to a state with a majority of active agents. The second problem is to find a disposition of loyalists that returns the network to a state with a majority of inactive agents. Similar problems are studied for networks in which simple agents demonstrate the contrary to conforming behavior that we call anticonforming. We obtained several theoretical results regarding the behavior of collectives of agents with conforming or anticonforming behavior. In computational experiments we solved the described problems for randomly generated networks with several hundred vertices. We reduced corresponding combinatorial problems to the Boolean satisfiability problem (SAT) and used modern SAT solvers to solve the instances obtained.

\section*{Introduction}

In recent years the interest to the analysis of various phenomena of collective behavior has significantly increased. It can be explained by the fact that in almost all areas of human activity there are processes involving information exchange inside collectives. Such processes deeply affect the future behavior of a collective and can lead to positive or negative consequences not only for the collective considered but also for a much larger social formation. For example an intensive sale of shares on the stock exchange market by players that have a big influence on others can lead to a drastic drop of global economic indexes. Riots and revolutionary situations proceed in a similar fashion when a relatively small group of instigators activates such a large number of people that state security systems are not able to cope with it.

The active development of social networking services in later years greatly increased the possibilities in collective behavior manipulation. This thesis can be proved by analyzing such revolutionary phenomena as Arab Spring, 2011-13 Russian protests, Euromaidan etc. In the majority of these cases the corresponding actions were planned via social networks. It is worth mentioning that such processes are usually coordinated by small groups of designated activists.

The modeling of collective behavior was studied in a large number of papers. Following many other authors we base our work on the paper of M. Granovetter \cite{Gran_1978}, which studied threshold models of collective behavior. The threshold behavior means that a state of every member of a group changes only when the value of a special function, that is associated with this agent, reaches some threshold. The simplest example of such behavior is to follow the decision of the majority. In Granovetters model the network connecting the agents is specified by a complete graph -- every agent takes into account the opinion of every other agent. In many real situations such approach cannot be used. For example, in real world social networks an agent usually bases its opinion on that of agents from some neighborhood. In this case the opinion of agents outside of such neighborhood would have no impact on the opinion of the agent considered. Similar situations can be observed in genetics: in many gene networks the amount of genes that directly affect each particular gene is small relative to the total number of genes in the network.

Similarities of dynamical processes that can be observed in gene networks and social networks led us to an idea to introduce and analyze models of collective behavior that are based on Boolean networks. The apparatus of Boolean networks have been used in mathematical biology for 50 years. Below we consider the so called synchronous Boolean networks (SBNs) first introduced by S. Kauffman in \cite{Kauffman1969} with the purpose of analyzing dynamical properties of gene networks.
In our approach we consider a collective as an SBN with special functions associated with the network vertices. From our point of view the language of Boolean networks is well suited for explaining a number of phenomena of collective behavior. For example, equilibrium states from \cite{Gran_1978} can be viewed as fixed points of a discrete function specified by the corresponding SBN. Another important feature of such models is that to solve combinatorial problems that arise during the analysis of SBNs it is possible to use modern methods of solving large systems of Boolean equations. For this purpose in our paper we use algorithms for solving the Boolean satisfiability problem (SAT).

Let us present a brief outline of the paper. First we describe SBNs and define fixed points and cycles of discrete functions determined by these networks. Then we introduce two models of collective behavior that are based on SBNs. In the first model we consider a situation when each network agent at the next moment of time makes a decision to act if at least a specific amount of agents in its neighborhood are currently active. Otherwise the agent decides not to act. This form of collective behavior is usually referred to as conformity. The second model is used to illustrate the phenomenon of anticonformity - an agent decides not to act if at the present moment at least a specific amount of its neighbors decide to act and vice versa. After this, we extend the models proposed by introducing two special types of agents: instigators and loyalists. Instigators are the agents that always act regardless of other agents decisions. Loyalists are the agents that never act. For the extended models we formulate the following combinatorial problem: for a network with a majority of inactive agents to find such a disposition of small amount of instigators, that after several moments of time the majority of agents in this network becomes active. An opposite problem is also considered: for a previously activated network (with instigators), to find a disposition of a relatively small number of loyalists, such that after several moments of time the majority of agents becomes inactive. In the context of problems considered we state a number of theoretical properties of discrete functions defined by the corresponding SBNs. Then we note that modern combinatorial algorithms can be used to solve such problems. In particular we use algorithms for solving SAT. Further we describe our computational experiments and discuss the results. In these experiments we constructed SBNs according to widely known models of random graphs (Gilbert-Erdos-Renyi model, Watts-Strogatz model, Barabasi-Albert model). Using modern SAT solvers we managed to solve combinatorial problems outlined above for corresponding networks with 500 vertices and more. In the conclusion we give some final remarks and outline our future plans.

\subsection*{Related Works}
As we already noted, the paper \cite{Gran_1978} is the fundamental work in the field of threshold models of collective behavior.
In a number of later works, for example \cite{Gran-Soong83,Gran-Soong86,Braun_1995}, the ideas from \cite{Gran_1978} were detailed and applied to analysis of various sociological situations.

In \cite{Chwe_1996,Breer_AT,novikov2013theory,novikov2014reflexion} and others it was shown that various phenomena of collective behavior may be studied from the game theory point of view. In particular, equilibrium states \cite{Gran_1978} in collectives can be considered as Nash equilibria. In this context we would like to mention the work \cite{Breer_AT} in which the conformity and anticonformity phenomena were considered from the game theory positions.

In the paper \cite{Chiang-2007} the influence of thresholds distributions on the genesis and development of several phenomena (in particular, the so called bandwagon effect) in the networks with arbitrary structure is analyzed.

As we said above, synchronous Boolean networks were introduced by S. Kauffman in \cite{Kauffman1969}. In that paper problems of analysis of fixed points and cycles of corresponding discrete functions were considered as important and helpful for the study of dynamics of real gene networks. Apparently, \cite{DBLP:conf/iccad/DubrovaTM05} is the first example of application of combinatorial algorithms to the search for cycles of discrete functions specified by Kauffman networks. Later the same authors used the SAT approach for similar purposes \cite{DubrovaT11}. In \cite{EKS} we considered the problem of search for fixed points of discrete functions specified by networks, in which vertex weight functions take natural values and at the same time act as threshold functions. In order to solve the corresponding problems, we used both SAT and ROBDD approaches. Also in \cite{EKS} we studied an opposite problem: given fixed points of the function specified by some network, to restore the structure of the network. 

In recent years there were published a lot of works about the analysis of structure of big networks and processes that can occur in them. Works \cite{Newman03thestructure} and \cite{Dorogovtsev2008} are quite complete reviews of relevant topics.


\section*{Models}
\subsection*{Synchronous Boolean Networks}

A \textit{Synchronous Boolean Network} (SBN) is defined as a directed graph in which with each vertex there is associated a total function that takes values from $\{0,1\}$ at discrete moments of time. Hereinafter we will refer to such functions as vertex \textit{weight functions}. The value of a weight function for an arbitrary vertex $v$ at moment $t+1$ is calculated based on the values of weight functions of some set of network vertices at moment $t$. In SBNs values of all weight functions are updated simultaneously (synchronously). Note that the weight functions can be specified in various ways: by truth tables, Boolean formulas or predicates. Values of weight functions of all vertices at an arbitrary moment $t$, $t\geq 1$ can be considered as a result of computing a value of a discrete function that takes a Boolean vector of length $n$ as input and outputs a Boolean vector of length $n$, where $n$ is a number of vertices in the network. We denote a Boolean vector consisting of weight functions values at moment $t$ as $W(t)$ and call it a \textit{network state} at moment $t$. We will refer to $W(0)$ as to an \textit{initial network state}. It is clear that an arbitrary SBN with $n$ vertices has $2^n$ different possible network states.

Thus, more formally, let us assume that $G$ is a directed graph with $n$ vertices that represents some SBN. Below we will consider only graphs without loops and without multiple arcs. For convenience let us mark vertices by natural numbers from $1$ to $n$. With an arbitrary vertex $v_i$, $i\in\{1,\ldots,n\}$ we associate a weight function $f_{v_i}(t)$, whose values are defined at discrete moments of time $t\in\{0,1,2,\ldots\}$. We assume that at $t=0$ each weight function has some initial value. By $V_i$ we denote such a set of network vertices that for each $v_j\in V_i$, $v_j\neq v_i$ the graph $G$ has an arc $(v_j, v_i)$. Essentially it means that $V_i$ contains vertices that directly affect $v_i$. We also call $V_i$ a \textit{neighborhood} of $v_i$. 

\begin{figure}[ht]
\centering
\includegraphics{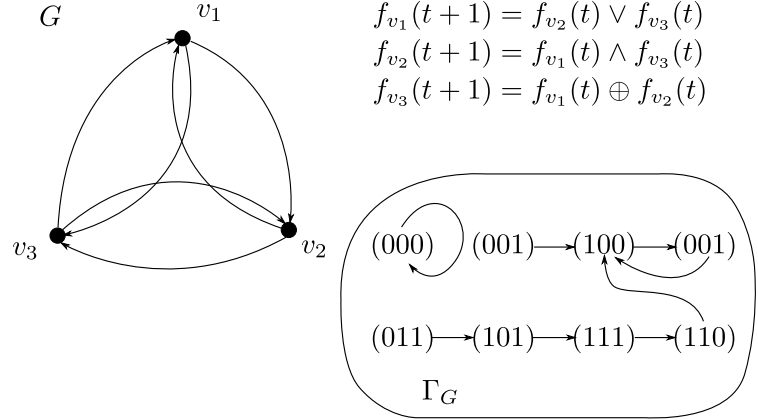}
\caption{
{\bf An example of a Kauffman network and its State Transition Graph.} The left part shows a simple Kauffman network with 3 vertices. Weight functions are specified by Boolean formulas in the right upper part of the figure. The lower right part demonstrates a state transition graph (STG) for the discrete function specified by this network. It contains one cycle of length 2 and one fixed point.
}
\label{kau_fig}
\end{figure}

From here on by $\{0,1\}^n$ we mean the set of all possible binary words of length $n$. The rules that specify each weight function $f_{v_i}$, $i\in\{1,\ldots,n\}$ are the same at any moment of time. It means that in total these functions specify a vector function that is defined everywhere in $\{0,1\}^n$ and takes values from $\{0,1\}^n$. We denote this function as $F_G:\{0,1\}^n\rightarrow\{0,1\}^n$ and refer to it as a \textit{discrete function defined by network} $G$. The transitions between network states, represented by Boolean vectors from $\{0,1\}^n$, can be naturally illustrated using special graphs called \textit{State Transition Graphs} (STGs). We denote the STG of network $G$ as $\Gamma_G$. An example of a simple SBN with 3 vertices where weight functions are specified by Boolean formulas is displayed in Figure \ref{kau_fig}.

As we already noted, the amount of different states of an arbitrary SBN with $n$ vertices is $2^n$, and the rules, according to which the network transitions from one state into another, do not depend on $t$. Therefore, regardless of the network state at moment $t=0$, there are such $k$ and $l$, $0\leq k<l$, that $W(k)=W(l)$. In this situation we call the sequence of transitions $W(k)\rightarrow\ldots\rightarrow W(l)$ a cycle of length $l-k$ \cite{Kauffman1969}. In some works on the analysis of dynamical properties of gene networks the cycles are called "attractors". The cycle of length 1 is called a fixed point of function $F_G$. For the network in Figure \ref{kau_fig} it is easy to see that $(000)\rightarrow (000)$ is a fixed point, while a sequence $(100)\rightarrow(001)\rightarrow(100)$ forms a cycle of length 2. Note that the neighborhood of every vertex of the network in Figure \ref{kau_fig} is formed by other two vertices.

\section* {Models of Collective Behavior Based on Synchronous Boolean Networks}
In this section we introduce and analyze two phenomena of collective behavior that can be observed in the real world. The first one is conforming behavior. It means that an agent agrees with the opinion of some agents from its neighborhood. It is easy to find many examples of conformity in real life: from riots and financial crises mentioned above to presidential elections, etc. The second phenomenon we study is anticonforming behavior. The agent demonstrating anticonforming behavior acts as an opposite to an agent with conforming behavior: it chooses not to act while certain amount of agents from its neighborhood are active and vice versa.

Let us consider an SBN $G$ with $n$ vertices interpreting agents. We will say that an arbitrary agent $v_i$, $i\in \{1,\ldots,n\}$ is \textit{active} (\textit{inactive}) at moment $t$ if $f_{v_i}(t)=1$ ($f_{v_i}(t)=0$, respectively). We assume that an arbitrary agent $v_i$ is associated with the weight function of one of the following two types:
\begin{equation}
\label{function_conf}
f_{v_i}(t+1)=\left\{
\begin{array}{c}
1, \sum\limits_{v_j\in V_i} f_{v_j}(t)\geq \theta_i\cdot |V_i|\\
0, \sum\limits_{v_j\in V_i} f_{v_j}(t)<\theta_i\cdot |V_i|
\end{array}
\right.
\end{equation}

\begin{equation}
\label{function_anticonf}
f_{v_i}(t+1)=\left\{
\begin{array}{c}
0, \sum\limits_{v_j\in V_i} f_{v_j}(t)\geq \zeta_i\cdot |V_i|\\
1, \sum\limits_{v_j\in V_i} f_{v_j}(t)<\zeta_i\cdot |V_i|
\end{array}
\right.
\end{equation}
where $\theta_i, \zeta_i \in [0,1]$ are called \textit{conformity threshold} and \textit{anticonformity treshold}, respectively. 

Essentially, \eqref{function_conf} means that the agent $v_i$ becomes active at moment $t+1$ only if at least $\lceil \theta_i \cdot |V_i|\rceil$ agents from its neighborhood are active at moment $t$. Otherwise $v_i$ becomes inactive at moment $t+1$. Hereinafter we refer to such agents as to \textit{conformists}. Likewise \eqref{function_anticonf} means that $v_i$ becomes inactive at moment $t+1$ if at least $\lceil \zeta_i \cdot |V_i|\rceil$ agents from its neighborhood are active at moment $t$ and becomes active otherwise. These agents will be refered to as \textit{anticonformists}. Values $\Theta_i=\lceil \theta_i\cdot |V_i|\rceil$ and $Z_i=\lceil \zeta_i\cdot|V_i|\rceil$ we will call \textit{conformity level} and \textit{anticonformity level}, respectively. Further we assume that if $V_i=\emptyset$ then the sum of corresponding weights is $0$.

Let $v_i$ be a conformist with the conformity threshold $\theta_i=0$ and $f_{v_i}(0)=1$. Then it is clear that $f_{v_i}(t)\equiv1$, i.e. that $f_{v_i}(t)$ takes the value of $1$ at any moment $t$. It means that agent $v_i$ is active at any moment regardless of decisions of agents in its neighborhood. We will refer to such agents  as to \textit{instigators}.

Now let $v_i$ be an anticonformist with anticonformity threshold $\zeta_i=0$ and $f_{v_i}(0)=0$. Following the similar reasoning we can conclude that such agent is inactive at any moment of time regardless of decisions of agents from its neighborhood. We call such agents \textit{loyalists}.

To an arbitrary agent that is neither instigator nor loyalist we will refer to as a \textit{simple agent}. Thus an arbitrary simple agent $v_i$ is either a conformist with $\theta_i>0$ or an anticonformist with $\zeta_i>0$.

\begin{figure}[htb]
\centering
\includegraphics{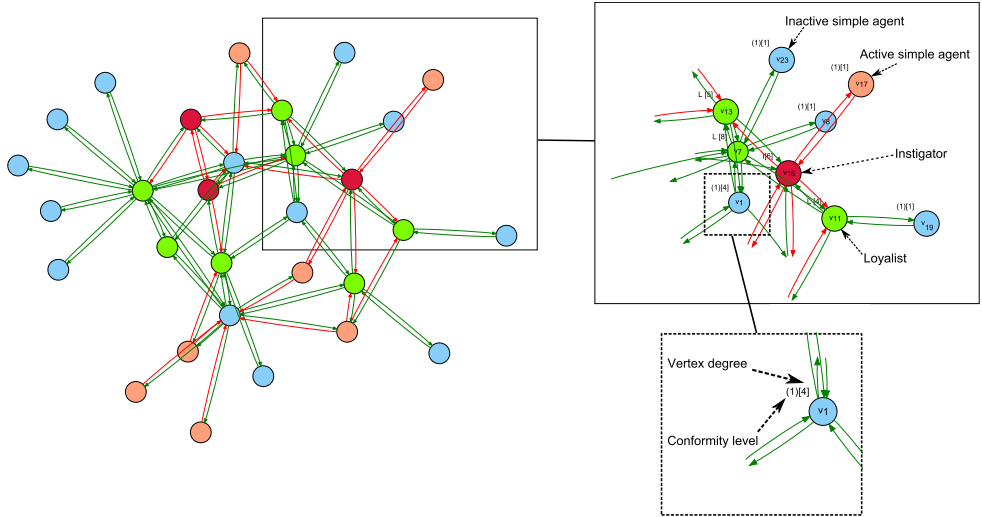}
\caption{
{\bf Example of an SBN representing a collective with conforming behavior.} This figure shows a network with different types of vertices. Each vertex represents a member of a collective (an agent). Crimson vertices correspond to instigators -- agents that are always active. Bright green vertices represent loyalists -- agents that are always inactive. The vertex corresponding to simple agent that is neither instigators nor loyalist is marked with orange if the agent is active and with blue otherwise. The arcs going from active agents (including instigators) are marked with red. The arcs going from inactive agents (including loyalists) are marked with green. Each simple agent has a conformity level. 
}
\label{legend}
\end{figure}

In Figure \ref{legend} we demonstrate the notation that we use below.

The networks with described types of agents can often be observed in real life. Indeed, for example one can notice that on the early stage of every revolutionary situation there are instigators. Their purpose is to activate as many initially inactive simple agents-conformists as possible. Once they become active, conformists help activating other inactive agents-conformists in the following moments of time. This process gradually involves even agents that are not directly connected to instigators. The goal of loyalists in such situations is to launch the deactivation process aimed at making active simple agents inactive.

\begin{figure}[htb]
\centering
\includegraphics{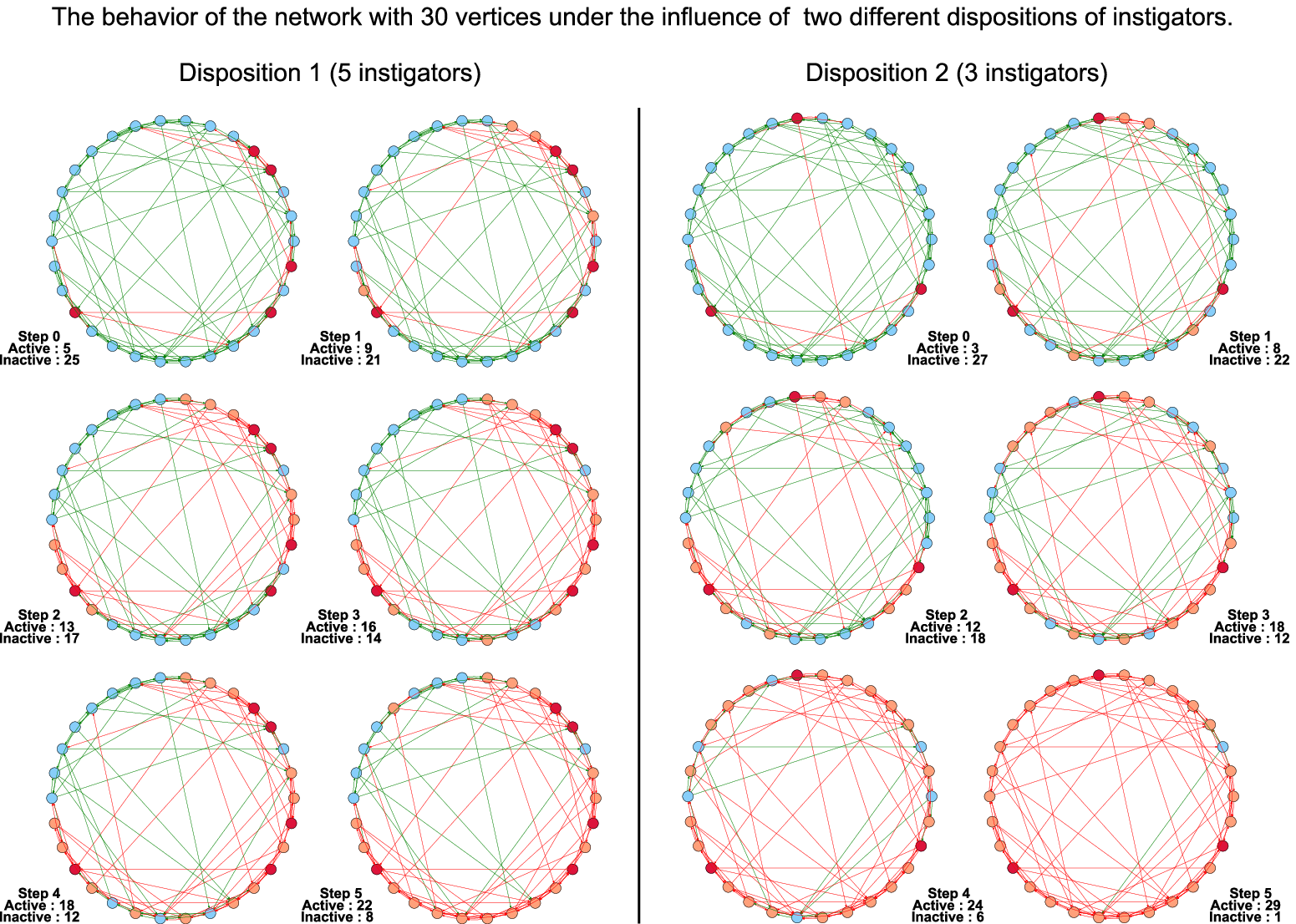}
\caption{
{\bf The behavioral dynamics of the network under the influence of two different dispositions of instigators.} In the initial state all simple agents are inactive. In the first case (left part of the figure), 5 instigators after 5 steps activate 17 simple agents. In the second case (right part of the figure) 3 instigators after 5 steps activate 26 simple agents.
}
\label{instigators_2vars}
\end{figure}

It should be noted that the disposition of instigators and loyalists in the network can significantly affect the activation / deactivation of the network. In Figure \ref{instigators_2vars} we display the behavior of one network with two different dispositions of instigators at the initial time moment. The considered networks does not have loyalists and all its simple agents are conformists. We assume that at the initial moment all the simple agents are inactive (i.e. for every simple agent $f_{v_i}(0)=0$). In the first case 5 instigators after 5 moments of time manage to activate only 17 simple agents. In the second case 3 instigators after 5 moments of time activate almost the whole network --- 26 simple agents. Important detail here is that in the first case there is more instigators but their disposition is worse.

Further we establish a number of theoretical results regarding the dynamical properties of SBNs with agents of the described types. The main achievement here consists in the justification of the fact that the networks in which all simple agents are conformists and networks where all simple agents are anticonformists can demonstrate completely different activation / deactivation dynamics.

\subsection*{Conforming Behavior}

Consider an arbitrary SBN $G$ with $n$ agents. We assume that all the simple agents in the network are conformists and that there can be instigators and loyalists. Hereinafter we study two problems that we believe to be interesting from the practical point of view.

In the context of the first problem (to which we will refer below as \textbf{Problem 1}) we consider a network with $n$ agents among which there can be $I$, $I<n$ instigators, while all the other $n-I$ agents are simple agents-conformists. We assume that a priori $I$ instigators can be arbitrarily placed in the network. Also we assume that at the initial time moment $t=0$ all the simple agents are inactive. The goal is to find the disposition of instigators such that starting from $t=0$ the network after some time moments  transitions to the state with a majority of active agents.

The second problem (to which we will refer below as \textbf{Problem 2}) consists in the following: we consider the network with a fixed disposition of $I$, $I<n$ instigators and all the other $n-I$ simple agents-conformists are active at the initial moment $t=0$. We assume that it is possible to replace $L$, $L< n-I$ arbitrary simple agents by loyalists. We need to find such disposition of these loyalists that  starting from $t=0$ the network after some time moments transitions to the state with the majority of inactive simple agents.

Let us show that the following theorem holds.

\textit{\textbf{Theorem 1.}
Consider an arbitrary SBN with $n$ agents among which there are $I$, $I<n$ instigators and the remaining $n-I$ simple agents are conformists. We assume that at the initial time moment $t=0$ all $n-I$ simple agents are inactive. Then for any disposition of instigators and any conformity thresholds of simple agents the network starting from $t=0$ will transition to a fixed point after $T\leq n-I$ time moments.}

\textbf{Proof.} Assume that $G$ is an SBN with $n$ vertices, weight functions \eqref{function_conf}, an arbitrary disposition of $I$ instigators and arbitrary conformity thresholds of simple agents. Suppose that all simple agents are inactive at $t=0$. If after transition from $t=0$ to $t=1$ none of simple agents have changed their decisions ($0\rightarrow 1$) then we have a fixed point (since instigators do not change their decisions by definition). Now suppose that at moment $t=1$ some simple agents have changed their decisions from $0$ to $1$. Let $v$ be one of them. It means that $v$ has changed its decision from $0$ to $1$ only because it had enough (relative to its conformity threshold) instigators in its neighborhood. But since instigators are always active then the number of active agents in the neighborhood of $v$ at any $t\geq 1$ can not be less than that at $t=0$. Therefore this agent will not change its decision $1$ at any of the following moments of time. If at moment $t=2$ none of simple agents have changed their decisions then we have a fixed point. Suppose $v$ is an arbitrary agent that has changed its decision during the transition from $t=1$ to $t=2$. From the above it follows that $v$ changed decision from $0$ to $1$. It could have occured only because it had enough (relative to its conformity threshold) instigators and active agents in its neighborhood. However all agents that have become active at $t=1$ cannot change their decisions at the following moments of time. Therefore agent $v$ will remain active at all $t\geq 2$. If we continue by analogy we can conclude that not later than after $n-I$ time moments our network will reach a fixed point. $\blacksquare$

Using the reasoning technique from the proof of Theorem 1 it is easy to prove the following corollary.

\textit{
\textbf{Corollary 1.}
Consider an arbitrary SBN with $n$ agents among which there are $I$, $I<n$ instigators and the remaining $n-I$ simple agents are conformists. Assume that some disposition of instigators is fixed and all simple agents are active at the initial time moment $t=0$. Also assume that we can replace any $L$, $L<n-I$ simple agents to loyalists. Then for any disposition of these $L$ loyalists and any conformity thresholds of remaining $n-I-L$ simple agents the network starting from $t=0$ will transition to a fixed point after $T\leq n-I-L$ time moments.
}

Note that the Theorem 1 despite its simplicity makes it possible to explain the situations when a relatively small number of instigators thanks to their advantageous disposition manage to activate quite a big network quickly. Apparently, the development of revolutionary situations, epidemics and critical processes in stock markets proceed in the similar fashion.

The principal possibility of the phenomenon when a small number of instigators can activate the network starting from the state in which all simple agents-conformists are inactive means that the network itself is vulnerable to instigators. Intuitively it is clear that such networks can be activated by instigators even faster if some simple agents are already active at the initial time moment. This thesis is proved by the following theorem.

\textit{\textbf{Theorem 2.}
Assume $S_0(I)$ is a state of an SBN with $n$ vertices with weight functions \eqref{function_conf} and $I$, $I<n$ instigators, in which all simple agents-conformists are inactive. Denote by $S(I)$ a network state, with the same disposition of instigators as in $S_0(I)$, in which there is at least one active simple agent. By $W_0(T)$ and $W(T)$ we denote states reached by the network after $T$ time moments starting from $S_0(I)$ and $S(I)$, respectively. Then for any $T\in N^{+}$
\begin{equation*}
wt(W_0(T))\leq wt(W(T)),
\end{equation*}
where $wt(x)$ stands for a Hamming weight of a Boolean vector $x$.
}

\textbf{Proof.} Consider a state $S_0(I)$ in which all simple agents are inactive and a state $S(I)$ where some $k$, $k\geq 1$ simple agents are active. Denote these active agents as $a_1,\ldots, a_k$. We assume that the disposition of $I$ instigators is the same in both $S_0(I)$ and $S(I)$. First let us prove that $wt(W_0(1))\leq wt(W(1))$. Let us analyze all possible cases. First, both $S_0(I)$ and $S(I)$ can be fixed points of $F_G$. In this case the property holds. If $S_0(I)$ is a fixed point and $S(I)$ is not, then even if all agents $a_1,\ldots, a_k$ become inactive in $W(1)$ it holds that $wt(W_0(1))\leq wt(W(1))$. Now suppose that $S_0(I)$ is not a fixed point, i.e. some simple agents in $W_0(1)$ become active. It can only occur if they have enough instigators in their neighborhoods (relative to their conformity thresholds). But it means that the same simple agents will be active in $W(1)$. Additionally some (possibly all) agents from $a_1,\ldots,a_k$ can become inactive or remain active in $W(1)$. Also in $W(1)$ there can appear other active simple agents because $a_1,\ldots,a_k$ are active in $S(I)$. In any case we have $wt(W_0(1))\leq wt(W(1))$. Since $S_0(I)$ is not a fixed point of $F_G$ then some simple agents in $W_0(1)$ become active. Denote these agents as $b_1,\ldots,b_s$. From Theorem 1 it follows that these agents cannot become inactive in any of states $W_0(1), W_0(2),\ldots$. Consider an arbitrary agent $b_i$, $i\in \{1,\ldots,s\}$ and let $V_{b_i}$ be its neighborhood. From the above the number of active agents in $V_{b_i}$ in states $W(1), W(2),\ldots$ is not less than that in $V_{b_i}$ in the state $S_0(I)$. Therefore $b_1,\ldots, b_s$ will be active in all states $W(r)$, $r\geq 1$. It means that $W_0(1)$ and $W(1)$ can be considered as initial states of the network with a set of $I+s$ instigators: this set is formed by $I$ original instigators and $s$ new instigators $b_1,\ldots, b_s$. After that by analogy we can show that $wt(W_0(2))\leq wt(W(2))$, etc. $\blacksquare$

\subsection* {Anticonforming Behavior}
Consider an arbitrary SBN $G$ with $n$ agents. We assume that all simple agents in $G$ are anticonformists and also that the network can contain instigators and loyalists.

On the first glance it may seem that the dynamical processes we studied for the collectives of conformists should have some simple analogues in the collectives of anticonformists. However, more thorough investigation reveals that this is not the case. In particular, assume that $G$ is a network in which any agent $v_i$ has a nonempty neighborhood ($V_i\neq \emptyset)$. Also let this network contain neither instigators nor loyalists. Then it is easy to see that if all the agents in the network are conformists (with non-zero conformity thresholds), then the states $0^n$ and $1^n$ are fixed points. However, if all the agents are anticonformists (with non-zero anticonformity thresholds) then there is the cycle of length $2$: $0^n\rightarrow 1^n \rightarrow 0^n$. Indeed, let $G$ be the network for which all listed conditions are satisfied, all its simple agents are anticonformists and they are inactive at moment $t=0$. Let $v_i$ be an arbitrary agent of the network and $V_i$ be its neighborhood. Since $V_i\neq \emptyset$ (by assumption), then at $t=0$ in $V_i$ all the agents have the $0$ state. Therefore for any value of $\zeta>0$ we have: $\sum\limits_{v_j\in V_i}f_{v_j}(0)<\zeta\cdot|V_i|$, so at moment $t=1$ the agent $v_i$ will switch its state to $1$. Since $v_i$ is an arbitrary network agent, it means that at moment $t=1$ every agent of the network will switch to the state $1$.
 Now let us consider what occurs at moment $t=2$. Let $v_i$ be an arbitrary agent-anticonformist. Then at moment $t=1$ all the agents in $V_i$ are in the state $1$. It means that for any $0<\zeta\leq 1$ the following holds: $\sum\limits_{v_j\in V_i} f_{v_j}(1)\geq \zeta\cdot|V_i|$. In this situation at moment $t=2$ the agent $v_i$ switches to the state $0$. But since $v_i$ is an arbitrary agent, then all the network agents switch to $0$ at $t=2$. Therefore we have the cycle $0^n\rightarrow 1^n \rightarrow 0^n$.

The following theorem describes the dynamics of collectives of anticonformists with the same initial conditions as in Theorem 1. It can be noted that in this situation, generally speaking, the collective of anticonformists has more complex behavior than that of the collective of conformists. In particular, if the network of anticonformists starts from an initial state in which all simple agents-anticonformists are inactive, then it may not reach an equilibrium state (a fixed point).

\textit{
\textbf{Theorem 3.}
Consider an arbitrary SBN with $n$ agents, where $I$, $I<n$ agents are instigators and the remaining $n-I$ simple agents are anticonformists. Assume that at the initial moment $t=0$ all $n-I$ simple agents are inactive. Then for any disposition of instigators and any anticonformity thresholds of simple agents the network starting from $t=0$ after $T\leq n-I$ will either transition to a fixed point or will enter the cycle of length $2$.
}

\textbf{Proof.} Let $G$ be an SBN with $n$ vertices, weight functions \eqref{function_anticonf}, an arbitrary disposition of $I$ instigators and arbitrary anticonformity thresholds of simple agents. Below we denote the set of all vertices of $G$ as $V$. Let $S_0(I)$ be an initial state of a network with an arbitrary disposition of $I$ instigators and with inactive simple agents. Let $W_0(1)$ be a state to which the network transitions from $S_0(I)$ at moment $t=1$. If in $W_0(1)$ none of simple agents have changed their decisions (from $0$ to $1$) then we have a fixed point. Suppose that $m=n-I$, $m>0$ and $r$, $0<r\leq m$ simple agents have switched from $0$ to $1$. If $r=m$, i.e. all simple agents have switched, then with the transition from $W_0(1)$ to $W_0(2)$ all these agents will switch back from $1$ to $0$ since in $W_0(1)$ each of them has a neighborhood consisting only of active agents. Therefore in this case we have the following cycle of length $2$: $S_0(I)\rightarrow W_0(1) \rightarrow S_0(I)$. Now suppose that $r<m$. Consider $q=m-r$, $q>0$ simple agents that have not switched from $0$ to $1$ with the transition from $S_0(I)$ to $W_0(1)$. It could have occured only if in their neighborhoods there were enough (relative to their anticonformity thresholds) instigators (which are always active). But since instigators do not change their decisions, then each of these $q$ agents will not switch from $0$ to $1$ at any of the following time moments. Denote by $R_1$, $|R_1|=r$ the set formed by all simple agents that have switched ($0\rightarrow 1$) at moment $t=1$. Note that every agent from $V\backslash R_1$ does not change its state from $0$ to $1$ at time moments $t$, $t\geq 1$. Further let us look only at the behavior of agents from $R_1$. Consider moment $t=2$. If none of agents from $R_1$ have switched ($1\rightarrow 0$) then we have a fixed point (since all agents from $V\backslash R_1$ do not change their decisions at any $t\geq 1)$. Suppose that $p$ agents from $R_1$, $0<p\leq r$ have switched at $t=2$ ($1\rightarrow 0$). It is clear that if $p=r$ (all agents from $R_1$ have switched) then we have a cycle of length $2$. Assume $p<r$, by $Q$, $Q\subset R_1$ denote the set of all $r-p$ agents that have not switched ($1\rightarrow 0$) at moment $t=2$. Consider an arbitrary agent $v\in Q$. This agent has not changed its decision ($1\rightarrow 0$) at $t=2$ only because at $t=1$ its neighborhood had enough inactive agents from $V$ (relative to $v$ anticonformity threshold). However these inactive agents could not belong to $R_1$ (since at $t=1$ all agents from $R_1$ are active). Therefore they must belong to $V\backslash R_1$. But as we noted above all such agents do not change their decisions from $0$ to $1$ at any of moments $t\geq 1$. It means that any agent $v\in Q$ will not change its decision at any of the following moments $t\geq 2$. The set containing $p$, $p<r$ simple agents that have switched at $t=2$ from $1$ to $0$ we denote by $R_2$ and further analyze only the behavior of agents from $R_2$. By analogy we note that each agent from $V\backslash R_2$ does not change its decision at $t\geq 2$, etc. Thus at most after $T=n-I$ time moments the network considered will either reach a fixed point or enter a cycle of length $2$. $\blacksquare$

The reasoning technique from the proof of the Theorem 3 can be used to study the behavior of collectives of anticonformists, in which there are both instigators and loyalists. In particular, for an arbitrary network of such kind one can show that starting from the initial state, in which either all simple agents-anticonformists are active or inactive, the network will transition to a fixed point or will enter the cycle of length $2$ after at most $n-I-L$ time moments, where $I$ stands for the amount of instigators and $L$ - for the amount of loyalists.

\subsection*{Final Remarks}

In this part we presented several theoretical results regarding the conforming and the anticonforming behavior. From our point of view these results explain a number of phenomena observed in the real world. In particular, fast activation of a large network by a relatively small number of instigators can be explained not only by the network structure (for example by its strong connectivity) or by small conformity thresholds but also by advantageous disposition of instigators. If there exists such disposition of small number of instigators that forces the network to transition from the state with inactive simple agents to the state with a majority of active agents then this network is vulnerable to instigators. To determine the degree of such vulnerability for some particular disposition of $I$ instigators it is sufficient to study the behavioral dynamics of the network for at most $n-I$ time moments. This fact is the assertion of the Theorem 1. Evidently, for many real-world networks the vulnerability to instigators is highly undesirable. On the other hand, as it follows from the Corollary 1, even if the network was already activated by instigators, but there is a solution of \textbf{Problem 2}, then, roughly speaking, the situation can be improved by transforming a number of simple agents to loyalists.

Theorem 3 shows that the activation dynamics of collectives of anticonformists can significantly differ from that of the collectives of conformists even for the similar initial conditions. Unfortunately, we could not obtain any analogues of Theorems 1 and 3 for collectives in which simple agents are represented by both conformists and anticonformists. In the section about the experiments we give an example when such network displays more complex behavior.

\section*{SAT Approach to the Study of SBN-based Models of Collective Behavior}

Note that in the real world the conforming behavior is spread much more than the anticonforming. On the other hand, the collectives of anticonformists demonstrate more complex behavioral dynamics compared to that of collectives of conformists. It follows from theorems 1 and 3. That is why in our computational experiments we studied the collectives of conformists and concentrated our attention on \textbf{Problem 1} and \textbf{Problem 2}, formulated above. We would like to point out the fact that the considered problems are combinatorial since they presume the analysis of many possible variants of dispositions of instigators and loyalists. We applied to \textbf{Problem 1} and \textbf{Problem 2} the algorithms that are used to solve the Boolean satisfiability problem (SAT). This choice is motivated by the fact, that modern SAT solving algorithms are very powerful computational methods that successfully cope with combinatorial problems from a wide spectrum of practical areas \cite{HandbookOfSAT2009}.

For an arbitrary Boolean formula the Boolean satisfiability problem (SAT) consists in answering a question if this formula is satisfiable, i.e. if there exists such an assignment to Boolean variables of this formula, that makes the formula true. This problem in the general case can be effectively (in polynomial time on the length of a binary encoding of the considered formula) reduced to the problem of deciding if a Boolean formula in a conjunctive normal form (CNF) is satisfiable. Taking this fact into account, below we consider SAT in the following formulation: for an arbitrary CNF $C$ over the set of Boolean variables $X$ we need to answer a question if $C$ is satisfiable, and if the answer is 'yes', to present a corresponding variable assignment that evaluates $C$ to $1$. This problem is NP-hard, therefore, it cannot be solved in polynomial time if $P\neq NP$. Nevertheless, SAT is very important in a practical sense because a lot of industrial problems can be effectively reduced to it and solved using modern algorithms developed during recent 15 years. 
Basic algorithmic constructions used in solving SAT and main directions of development and applications of SAT approach are described in \cite{HandbookOfSAT2009}.

The reducibility of an arbitrary NP problem to SAT (in the form of decision problem) follows from the Cook theorem \cite{cook1971}. However, in practice the analysis of specific details of the considered problem makes it possible to significantly decrease the size of the CNF formula produced. A number of general techniques used to reduce combinatorial problems to SAT can be found in \cite{Pre09HBSAT}.

The SAT approach was successfully applied to the search for cycles of functions defined by Boolean networks in \cite{DubrovaT11} and \cite{Guo-2014}. It should be noted, however, that networks studied in that papers have their own specifics motivated by the source of origin: essentially they are Kauffman networks in which the power of the neighborhood of an arbitrary agent does not exceed some relatively small number $K$ (usually $K\in\{1,2,3\}$). Also, weight functions used in \cite{DubrovaT11} and \cite{Guo-2014} are completely different from the ones we use. That is why below we present a relatively detailed description of the SAT encoding process for problems outlined above.

Basic idea that is used to encode many combinatorial problems to SAT, including problems studied in our paper, is to represent the computation process for the considered discrete function (in our case it is $F_G:\{0,1\}^n\rightarrow \{0,1\}^n$) as a Boolean circuit $B(F_G)$ formed by logical gates from a complete basis (for example $\{\wedge,\neg\}$). Formally, circuit $B(F_G)$ is a directed acyclic graph where $n$ nodes are labeled as inputs. All other nodes of this graph are called inner nodes. Each inner node corresponds to logical gate from the chosen basis. Usually, nodes that form the output of the considered function are referred to as output gates. In our case circuit $B(F_G)$ has $n$ output gates.

Circuit inputs are labeled by Boolean variables $x_1,\ldots,x_n$. Below we refer to these variables as input variables. An output of each logical gate $E$ is marked by an auxiliary variable $u(E)$. By $\{y_1,\ldots,y_n\}$ we denote a set of $n$ variables corresponding to output gates. We refer to $\{y_1,\ldots,y_n\}$ as output variables. Let $U$ be the set of all auxiliary variables. Then $\{x_1,\ldots,x_n\}\cap U=\emptyset$, $\{y_1,\ldots,y_n\}\subseteq U$. For circuit $B(F_G)$ it is possible to effectively construct (in linear time on the total number of nodes in the circuit) a CNF $C(F_G)$. The corresponding procedure is based on the Tseitin transformations \cite{Tseitin83}.

Assume $E$ is an arbitrary gate in $B(F_G)$. If $E$ is a NOT-gate then it has a single input labeled by variable $p$. Then for NOT-gate $E$ we construct a formula $u(E)\leftrightarrow\neg p$ where by $\leftrightarrow$ we mean logical equivalence. CNF-representation of a Boolean function specified by formula $u(E)\leftrightarrow \neg p$ is 
\begin{equation*}
(u(E)\vee p)\wedge (\neg u(E)\vee \neg p)
\end{equation*}
If $E$ is an AND-gate, and $p,q$ are variables corresponding to its inputs, then for $E$ we construct formula $u(E)\leftrightarrow p\wedge q$ and CNF
\begin{equation*}
(\neg u(E)\vee p)\wedge (\neg u(E) \vee q)\wedge (u(E)\vee \neg p \vee \neg q)
\end{equation*}

We say that CNFs constructed this way encode the corresponding logical gates. Then the CNF encoding circuit $B(F_G)$ is
\begin{equation*}
C(F_G)=\bigwedge_{E\in B(F_G)}C(E)
\end{equation*}
where $C(E)$ is a CNF that encodes gate $E$.

Once we have a CNF $C(F_G)$ we can extend it by adding new constraints in the clausal form that specify function $F_G$ properties we are interested in. For example, a CNF
\begin{equation*}
C'(F_G)=C(F_G)\wedge C(x_1\leftrightarrow y_1)\wedge \ldots \wedge C(x_n\leftrightarrow y_n)
\end{equation*}
in which $C(x_i\leftrightarrow y_i)=(x_i\vee \neg y_i)\wedge (\neg x_i \vee y_i)$, $i\in\{1,\ldots,n\}$ specifies a fixed point of function $F_G$. To be more precise, CNF $C'(F_G)$ is satisfiable if and only if function $F_G$ has fixed points. If $C'(F_G)$ is satisfiable and its satisfying assignment is obtained, then we can effectively extract the corresponding fixed point: it is sufficient to write down values of the input variables. To make a SAT instance that specifies the problem of finding a cycle of length $k$ we need to represent a superposition 
\begin{equation*}
F_G^k=F_G\circ\ldots\circ F_G
\end{equation*}
as a Boolean circuit. 

Instead of logical gates we actually can use more complex basic Boolean functions, such as predicates over finite sets. In this case elements of the corresponding sets are represented by Boolean vectors. In fact this is what we do to encode functions $F_G:\{0,1\}^n\rightarrow \{0,1\}^n$ for networks with weight functions \eqref{function_conf} and \eqref{function_anticonf}.

Now let us consider an SBN with $n$ vertices and weight functions \eqref{function_conf} that can have both instigators and loyalists. Assume that the network is functioning for $T$ time moments. The decision of agent $v_i$, $i\in\{1,\ldots,n\}$ at moment $t\in\{0,1,\ldots,T\}$ we encode with Boolean variable $x_i^t$.

We would like to stress out once more that a priori we do not know dispositions of instigators and loyalists in the network and therefore presume that any agent can take one of these roles. To take into account that an arbitrary vertex $v_i$ can be either an instigator, a loyalist or a simple agent, we introduce two additional sets of Boolean variables $\{a_i\}_{i=1}^n$, $\{l_i\}_{i=1}^n$. We assume that if $a_i=1$, $l_i=0$ then $v_i$ is an instigator; if $a_i=0$, $l_i=1$, then it takes the role of a loyalist; if $a_i=l_i=0$ then our vertex represents a simple agent. The situation corresponding to $a_i=l_i=1$ would mean that the vertex is simultaneously an instigator and a loyalist. That is why it is forbidden by means of a clause $(\neg a_i\vee \neg l_i)$.

Let $v_i$ be an arbitrary network vertex, $V_i=\{v_{j_1},\ldots,v_{j_{|V_i|}}\}$ and $\Theta_i$ is conformity level of $v_i$. We introduce the following predicate
\begin{equation}
\label{big_predicate}
P_{\Theta_i}\left(x_{j_1}^t,\ldots,x^t_{j_{|V_i|}}\right)=
\left\{
\begin{array}{l}
\textit{True\;(1)}, \textit{ if } \sum\limits_{j\in\left\{j_1,\ldots,j_{|V_i|}\right\}} x_j^t\geq \Theta_i\\
\textit{False\;(0)}, \textit{ if } \sum\limits_{j\in\left\{j_1,\ldots,j_{|V_i|}\right\}} x_j^t< \Theta_i\\
\end{array}
\right.
\end{equation}

Then from the above we can conclude that the decision of agent $v_i$ at moment $t+1$ is associated with the following formula:
\begin{equation}
\label{eq3}
\left(x_i^{t+1}\leftrightarrow \neg l_i \wedge \left(a_i\vee P_{\Theta_i}\left(x^t_{j_1},\ldots,x^t_{j_{|V_i|}}\right)\right)\right)\wedge \left(\neg a_i\vee \neg l_i\right)
\end{equation}
Additional constraints on the initial network state are encoded in a similar fashion. For example a constraint that specifies that an arbitrary agent $v_i$ at the initial state is active only if it is an instigator is equivalent to satisfiability of the following formula:
\begin{equation}
\label{eq4}
\left(x_i^0\leftrightarrow a_i\right)\wedge \left(\neg a_i\vee \neg l_i\right)
\end{equation}
In fact, all clauses of the kind $\left(\neg a_i\vee \neg l_i\right)$ are added to the result CNF only once.

By applying Tseitin transformations to formulas \eqref{eq3} and \eqref{eq4} we can produce CNFs that are satisfiable if and only if the original Boolean formulas are satisfiable. To do this we need to be able to effectively encode predicate \eqref{big_predicate}. It can be represented as a Boolean circuit implementing a function that counts ones in a Boolean vector and then compares the obtained result with $\Theta_i$. Such circuit can then be encoded to CNF in accordance with the procedure described above. However, there are algorithms that produce more effective SAT encodings for predicates \eqref{big_predicate}. These algorithms are based on various methods that work with so called cardinality constraints (\cite{Een-SAT,Sinz05,DBLP:conf/cp/SilvaL07,Bailleux-2003,AsinNOR11}). In the present paper we encode predicates \eqref{big_predicate} using sorting networks. The main idea of the corresponding approach is very simple: we can sort bits in an arbitrary Boolean vector $\left(\alpha_1,\ldots,\alpha_m\right)$ descending from left to right, as we consider them as natural numbers from the set $\{0,1\}$. Let $\left(\beta_1,\ldots,\beta_m\right)$ be a result of such sorting. Then it is clear that $\sum\limits_{j=1}^m \alpha_j\geq k$, $k\in\{1,\ldots,m\}$ if and only if $\beta_k=1$. Essentially, in our work to sort Boolean vectors we used binary variants of Batcher sorting networks \cite{DBLP:conf/afips/Batcher68,cormen2009}. SAT encoding of such network with input $\left(\alpha_1,\ldots,\alpha_m\right)$ and output $\left(\beta_1,\ldots, \beta_m\right)$ requires $O(m\cdot log^2 m)$ auxiliary variables and $O(m\cdot log^2 m)$ clauses. SAT encodings for the constraints that specify that after $T$ time moments the network must contain at least $m$, $m\leq n$ active agents and the constraints of the kind $wt(a_1,\ldots,a_n)\leq I$, $wt(l_1,\ldots,l_n)\leq L$ are produced in a similar way.

It is easy to see that in the general case, if we encode the evolution of network $G$ with $n$ vertices during $T$ moments of time, then in the CNF obtained the number of variables and clauses will be upper-bounded by $O\left(T\cdot n^2\cdot log^2 n\right)$. Taking into account the theorems proved above for the combinatorial problems considered we can study only cases when $T\leq n-I-L$. 

We would like to briefly mention algorithms underlying the solvers that we have used to study the proposed models. As we said above, the book \cite{HandbookOfSAT2009} is probably the most complete source of information about the algorithms for solving SAT. There are several classes of such algorithms and their effectiveness is justified by their ability to solve real practical problems. To solve SAT instances encoding the combinatorial problems outlined above we used modern CDCL solvers, basic design features of which are described in \cite{MSLM09HBSAT}. This choice is motivated first by the fact that CDCL solvers provide us with exact solutions, and, second, these particular algorithms successfully cope with many hard SAT instances, for example, with instances that encode some cryptanalysis problems.

\section*{Results \& Discussion}

\subsection*{Computational Experiments}

In our computational experiments we constructed networks according to the known models of random graphs. In particular we used the Gilbert model \cite{Gilbert1959} also known as the Erdos-Renyi model \cite{erdos1959} (see also \cite{Solomonoff-1951}), the Watts-Strogatz model \cite{Watts-Colective-1998} and the Barabasi-Albert model \cite{Barabasi-1999}.

Informally the process of constructing tests for combinatorial problems outlined above for SBNs in which simple agents are conformists (tests for networks of anticonformists are generated in a similar way) looks as follows.
\begin{enumerate}
\item We generate a random oriented simple graph (without loops and without multiple arcs) with $n$ vertices, in the form of adjacency matrix where main diagonal is filled with zeros.
\item For each of $n$ vertices we generate a conformity threshold that is randomly selected from $[0,1]$ according to the uniform distribution.
\item For a fixed number of time moments $T$ we encode to SAT the problem of search for a disposition of instigators with given constraints on their number.
\item The CNF obtained is given to a SAT solver.
\item If the SAT solver managed to solve the instance, before exceeding the time limit, and found a satisfying assignment, then for the instigators disposition obtained we encode a problem of search for a disposition of loyalists with given constraint on their number.
\item The CNF obtained on the previous step is given to a SAT solver.
\item If the SAT solver managed to solve the provided instance and found a satisfying assignment then a corresponding disposition of loyalists is extracted.
\end{enumerate}

Now let us briefly describe random graph models that we used. In fact, original models generate undirected graphs, so we modified them to take into account all features of formulas \eqref{function_conf} and \eqref{function_anticonf} (the neighborhood $V_i$ of vertex $v_i$ is formed by vertices in $G$ that have arcs going to $v_i$).

When generating a graph according to the Gilbert-Erdyos-Renyi model we fix a parameter $p\in [0,1]$ that is a probability of an arc. Then an arbitrary element $g_{ij}$, $i\neq j$ of an adjacency matrix of graph $G$ takes the value of $1$ with probability $p$ and the value of $0$ with probability $1-p$.

An important feature of the original Watts-Strogatz model is that random graphs generated according to this model have the small-world property that can often be observed in real world networks. The parameters of the Watts-Strogatz model include $k$, $k\geq 2$ and $\beta\in [0,1]$. First we generate a regular lattice network with $n$ vertices, where each vertex $v_i$, $i\in \{1,\ldots,n\}$ is connected with an arc $\left(v_j,v_i\right)$ with $\frac{k}{2}$ vertices on either side of $v_i$ if $k$ is even. If $k$ is odd then we can consider $\lfloor \frac{k}{2}\rfloor$ and $\lceil \frac{k}{2} \rceil$ similar arcs $\left(v_j,v_i\right)$. On the second stage of graph generation each arc $\left(v_j,v_i\right)$ with probability $\beta$ is rewired to $\left(v_s,v_i\right)$, where $s$ is chosen according to the uniform distribution from some subset of $\{1,\ldots,n\}$ in such a way that in the resulting graph there will be no loops and no multiple arcs.

The Barabasi-Albert model is important because it allows one to generate random networks with scale-free property. The construction of a network according to the Barabasi-Albert model can be considered as an iterative process consisting of $S+1$ steps. On the step $s=0$ an initial network $G_0$ with $m_0$ vertices is built. The result of each step $s\in\{1,\ldots,S\}$ is the network $G_s$ which is constructed by adding to $G_{s-1}$ one new vertex $v'$ connected to $m\leq m_0$ existing vertices of $G_{s-1}$. The procedure of constructing edges $(v,v')$, $v\in G_{s-1}$ is probabilistic and is referred to as \textit{preferential attachment}. According to this procedure for $v'$ and an arbitrary $v\in G_{s-1}$ the edge $(v,v')$ is added to $G_s$ with probability
\begin{equation*}
\mathrm{Pr}\left(v',v\right)=\frac{\mathrm{deg}\; v}{\sum\limits_{\tilde{v}\in G_{s-1}}\mathrm{deg}\;\tilde{v}}
\end{equation*}
Step $s\in\{1,\ldots,S\}$ lasts, i.e. the corresponding probabilistic experiments are repeated, until vertex $v'$ is connected with $m$ vertices of the graph $G_{s-1}$. In our experiments we use the following modification of the Barabasi-Albert model. An open cycle, i.e. a cycle in which an edge connecting the first and the last vertices is removed, is used as an initial network $G_0$. On each step $s\in\{1,\ldots,S\}$ the probabilistic experiment is carried out for all pairs of the kind $(v',v)$ where $v\in G_{s-1}$, and as a result of the step new vertex $v'$ is connected with $\geq m$ existing vertices. In the final network every edge $\left(v',v\right)$ is replaced by a pair of arcs $\left(v',v\right)$ and $\left(v,v'\right)$. 

Defining the conformity thresholds of agents in real networks is a highly nontrivial task and in each particular case it requires a thorough analysis of the corresponding specifics. Since the main goal of our computational experiments was to test the general applicability of the SAT approach to the study of the considered models, we chose conformity thresholds for each vertex randomly.

In the series of experiments we considered networks with 500 vertices.  SAT instances were solved using the Plingeling SAT solver \cite{website:lingeling} working on 32 threads (two 16-core AMD Opteron 6276 CPUs with 64 GB RAM). The corresponding results are shown in tables \ref{tab:BA-results}, \ref{tab:WS-results} and \ref{tab:GNP-results}.

\begin{figure}[htbp]
\centering
\includegraphics{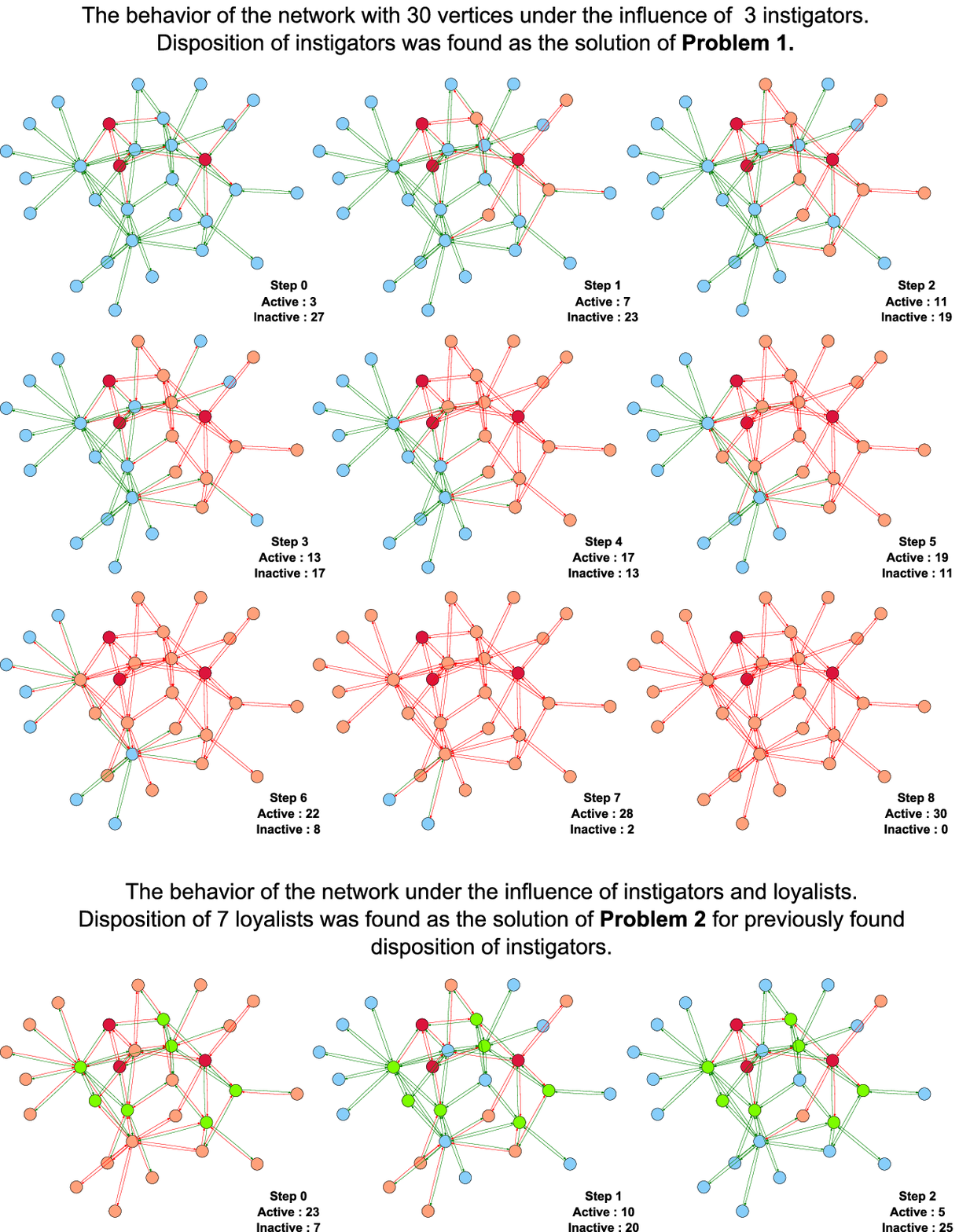}
\caption{
{\bf The behavior of the Barabasi-Albert network with 30 vertices under the influence of instigators and loyalists.} 
In the upper part of the figure the functioning of the network under the influence of 3 instigators is shown. In the lower part of the figure the functioning of the network under the influence of 3 instigators and 7 loyalists is shown. 
Dispositions of instigators and loyalists were found as solutions of \textbf{Problem 1} and \textbf{Problem 2}.
}
\label{bar30}
\end{figure}

\begin{figure}[htbp]
\centering
\includegraphics{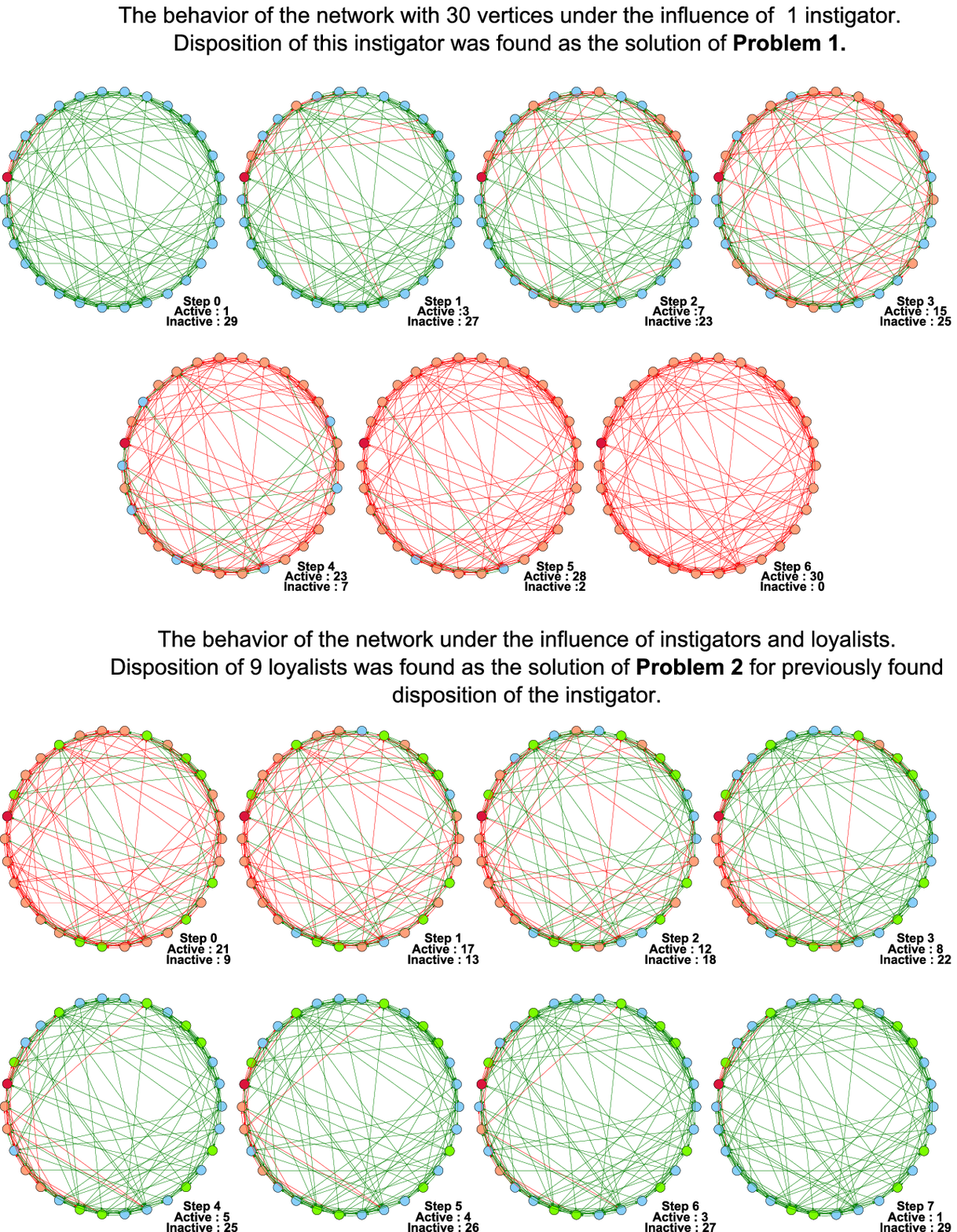}
\caption{
{\bf The behavior of the Watts-Strogatz network with 30 vertices under the influence of instigators and loyalists.} 
In the upper part of the figure the functioning of the network under the influence of 1 instigator is shown. In the lower part of the figure the functioning of the network under the influence of 1 instigator and 9 loyalists is shown. 
Dispositions of instigators and loyalists were found as solutions of \textbf{Problem 1} and \textbf{Problem 2}.
}
\label{ws30}
\end{figure}

Below we demonstrate several figures that illustrate the dynamics of SBNs with 30 vertices modeling the conforming behavior under the influence of instigators and loyalists. In Figure \ref{bar30} the evolution of the network generated according to the Barabasi-Albert model is displayed. In Figure \ref{ws30} we show that some networks (the particular network displayed was generated in accordance with the Watts-Strogatz model) are highly vulnerable to the influence of instigators. For the network shown it is sufficient to place one instigator to activate the whole network in $6$ steps. However, it is possible to find such disposition of $9$ loyalists that transforms the network to a state with the majority of inactive agents. 

\begin{figure}[htbp]
\centering
\includegraphics{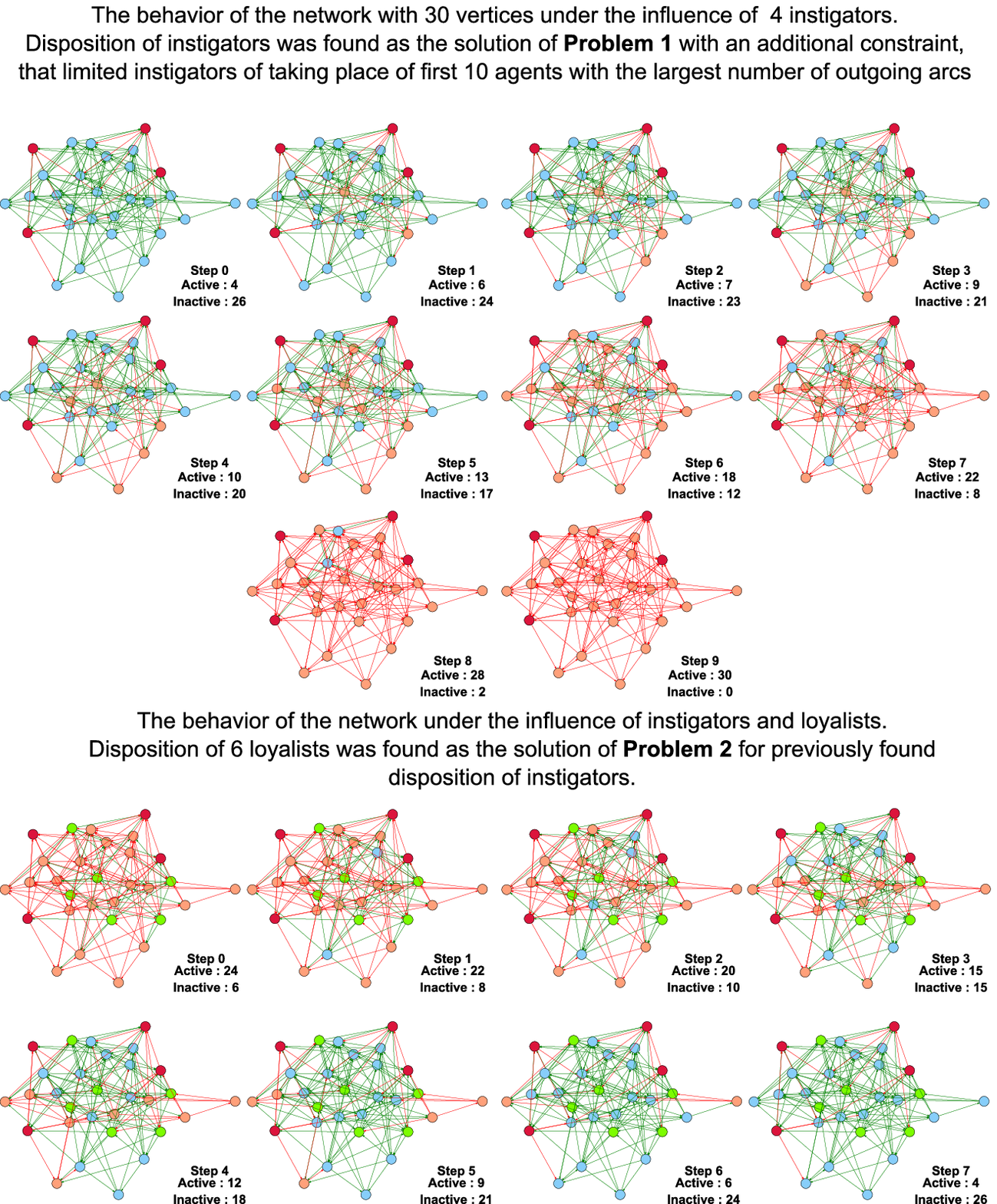}
\caption{
{\bf The behavior of the Erdos-Renyi network with 30 vertices under the influence of instigators and loyalists.} 
In the upper part of the figure the functioning of the network under the influence of 4 instigators is shown. In the lower part of the figure the functioning of the network under the influence of 4 instigators and 6 loyalists is shown. 
Dispositions of instigators and loyalists were found as solutions of \textbf{Problem 1} and \textbf{Problem 2}. Instigators could not take place of top 10 vertices with the largest number of outgoing arcs.
}
\label{gnp30}
\end{figure}

Intuitively, one of the most natural strategies of constructing dispositions of instigators is to place them into vertices with the largest number of outgoing arcs. In Figure \ref{gnp30} (the network is generated according to the Erdos-Renyi model) we show, that even if we forbid instigators to replace agents with the most advantageous positions (in the sense explained above), that does not exclude the existence of other possible variants of dispositions of instigators that transform the network into states with the majority of active agents. The corresponding constraints that forbid instigators and loyalists to take place of particular vertices are quite easily encoded into SAT.

Also we considered optimization variants of \textbf{Problem1} and \textbf{Problem2}, i.e. to find corresponding dispositions of instigators and loyalists of a minimal cardinality. These problems can also be effectively reduced to SAT using techniques described above. On the current stage we managed to solve corresponding problems for networks with 100 -- 150 vertices. 

\begin{table}[!ht]
\caption{
\bf{Results of the computational experiments for Barabasi-Albert networks with 500 vertices}}
\begin{tabular}{|c|c|c|c|c|}
\hline
$m$&\textbf{Pr1} CNF size, Kb& \textbf{Pr1} solving time, sec.& \textbf{Pr2} CNF size, Kb & \textbf{Pr2} solving time, sec.\\
\hline
0&13911.9&31.46&14350.9&1.97\\
2&22514.6&8.61&22957,4&3.44\\
4&51694.2&15.81&52187.1&168.73\\
8&134728.8&57.11&135232.6&342.43\\
\hline
\end{tabular}
\begin{flushleft}Results of the computational experiments for Barabasi-Albert networks, averaged for 10 tests (for each value of parameter $m$). \textbf{Pr1} and \textbf{Pr2} stand for \textbf{Problems 1} and \textbf{2} of finding dispositions of at most 50 instigators and at most 100 loyalists, respectively.
\end{flushleft}
\label{tab:BA-results}
\end{table}

\begin{table}[!ht]
\caption{
\bf{Results of the computational experiments for Watts-Strogatz networks with 500 vertices}}
\begin{tabular}{|c|c|c|c|c|c|}
\hline
$k$&$\beta$&\textbf{Pr1} CNF size, Kb& \textbf{Pr1} solving time, sec.&\textbf{Pr2} CNF size, Kb & \textbf{Pr2} solving time, sec.\\
\hline
10&0.2&53531.1&148.34&54023.1&811.55\\
10&0.3&51997.7&26.79&52490.8&3098.48\\
10&0.4&50891.1&16.51&51387.4&172.37\\
\hline
\end{tabular}
\begin{flushleft}Results of the computational experiments for Watts-Strogatz networks averaged for 10 tests (for each combination of values of parameters $k$ and $\beta$). \textbf{Pr1} and \textbf{Pr2} stand for \textbf{Problems 1} and \textbf{2} of finding dispositions of at most 50 instigators and at most 100 loyalists, respectively.
\end{flushleft}
\label{tab:WS-results}
\end{table}

\begin{table}[!ht]
\caption{
\bf{Results of the computational experiments for Erdos-Renyi networks with 500 vertices}}
\begin{tabular}{|c|c|c|c|c|}
\hline
$p$&\textbf{Pr1} CNF size, Kb& \textbf{Pr1} solving time, sec.& \textbf{Pr2} CNF size, Kb & \textbf{Pr2} solving time, sec.\\
\hline
0.01&17983.2&5.63&18425.5&46.69\\
0.02&51423.8&14.79&51918.6&16.74\\
0.03&105791.8&25.2&106293.8&34.49\\
\hline
\end{tabular}
\begin{flushleft}Results of the computational experiments for Erdos-Renyi networks, averaged for 10 tests (for each value of parameter $p$). \textbf{Pr1} and\textbf{ Pr2} stand for \textbf{Problems 1} and \textbf{2} of finding dispositions of at most 50 instigators and at most 100 loyalists, respectively.
\end{flushleft}
\label{tab:GNP-results}
\end{table}

In tables \ref{tab:BA-results}, \ref{tab:WS-results} and \ref{tab:GNP-results} we present the information about the size of encodings and about the time required to solve \textbf{Problems 1} and \textbf{2} on determining dispositions of instigators or loyalists. We considered networks with 500 vertices. For each value of parameter $p$ in case of Erdos-Renyi networks, combination of values of $\beta$ and $k$ in case of Watts-Strogatz networks, and $m$ in case of Barabasi-Albert networks we generated 10 different tests. Note, that solving time can greatly vary even within one test series (for a particular random graph model). From our point of view it can be explained by the fact that sometimes randomly generated tests are very hard for the particular SAT solver because of heuristics used, while the majority of such tests are solved relatively fast.

\subsection*{Conclusions and Future Works}
In the present paper we introduce the models of collective behavior, that are based on the synchronous Boolean networks, and study several phenomena related to conformity and anticonformity. In the context of the proposed models we formulate several combinatorial problems related to the search for dispositions of agents with special properties (instigators and loyalists) in a network. To these combinatorial problems we applied modern algorithms for solving the Boolean satisfiability problem (SAT). 

We do not pretend that the results of our paper can be directly applied to practice since all computational experiments were performed for artificially generated networks with a random structure. However, our main goal was to show the principal possibility of solving corresponding combinatorial problems for networks with hundreds of vertices. 

\begin{figure}[htb]
\centering
\includegraphics{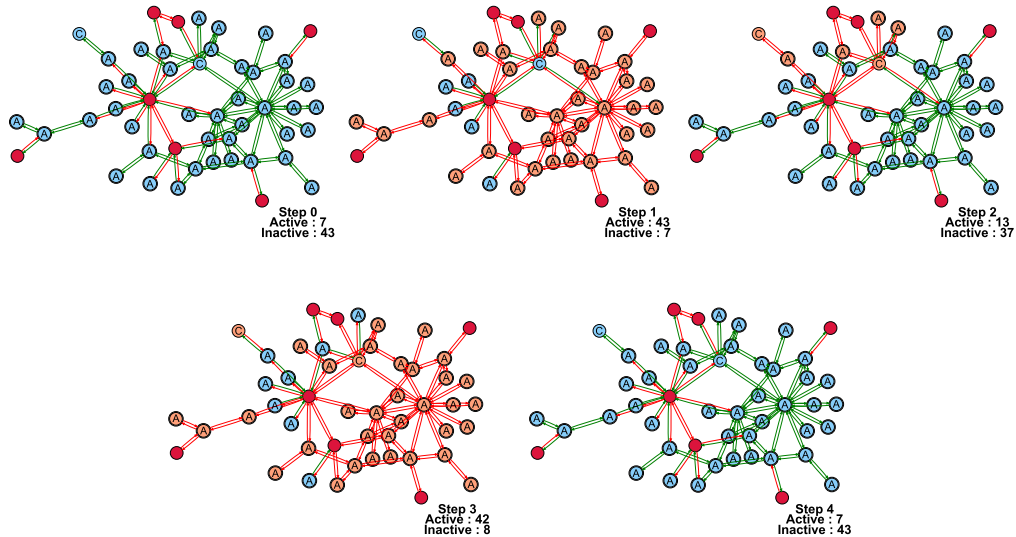}
\caption{
{\bf The cycle of length 4 for the network with both conformists and anticonformists.} The agents-conformists are marked with "C" and agents-anticonformists are marked with "A". The network contains 7 instigators (crimson vertices). At the initial time moment all simple agents are inactive.
}
\label{mixedcycle}
\end{figure}

We believe that the use of various SAT parallelization techniques will make it possible to develop our approach in such a way that it will be applicable to networks with 1000 and more vertices. The corresponding methods will be useful in the study of networks that represent strongly connected components extracted from the real world networks with a much greater number of vertices. The vulnerability of such strongly connected components to instigators in our opinion can have highly undesirable consequences for the corresponding large networks. To extract strongly connected components from real world networks, one can use methods from \cite{Vitali-2011}.

As we mentioned above, determining correct thresholds is probably the hardest stage of construction of any collective behavior model. In our experiments we generated such thresholds randomly. To study real world processes this task should be performed by a specialist in a relevant field of science (such as economy, biology, sociology, psychology, etc.).

Unfortunately we could not obtain the results similar to theorems 1 and 3 for the networks, in which simple agents are represented both by conformists and anticonformists. In Figure \ref{mixedcycle} we show how such network starting from the state in which all simple agents are inactive enters the cycle of length 4. It means that these networks display more complex behavior than that described by theorems 1 and 3.

\begin{figure}[htb]
\centering
\includegraphics{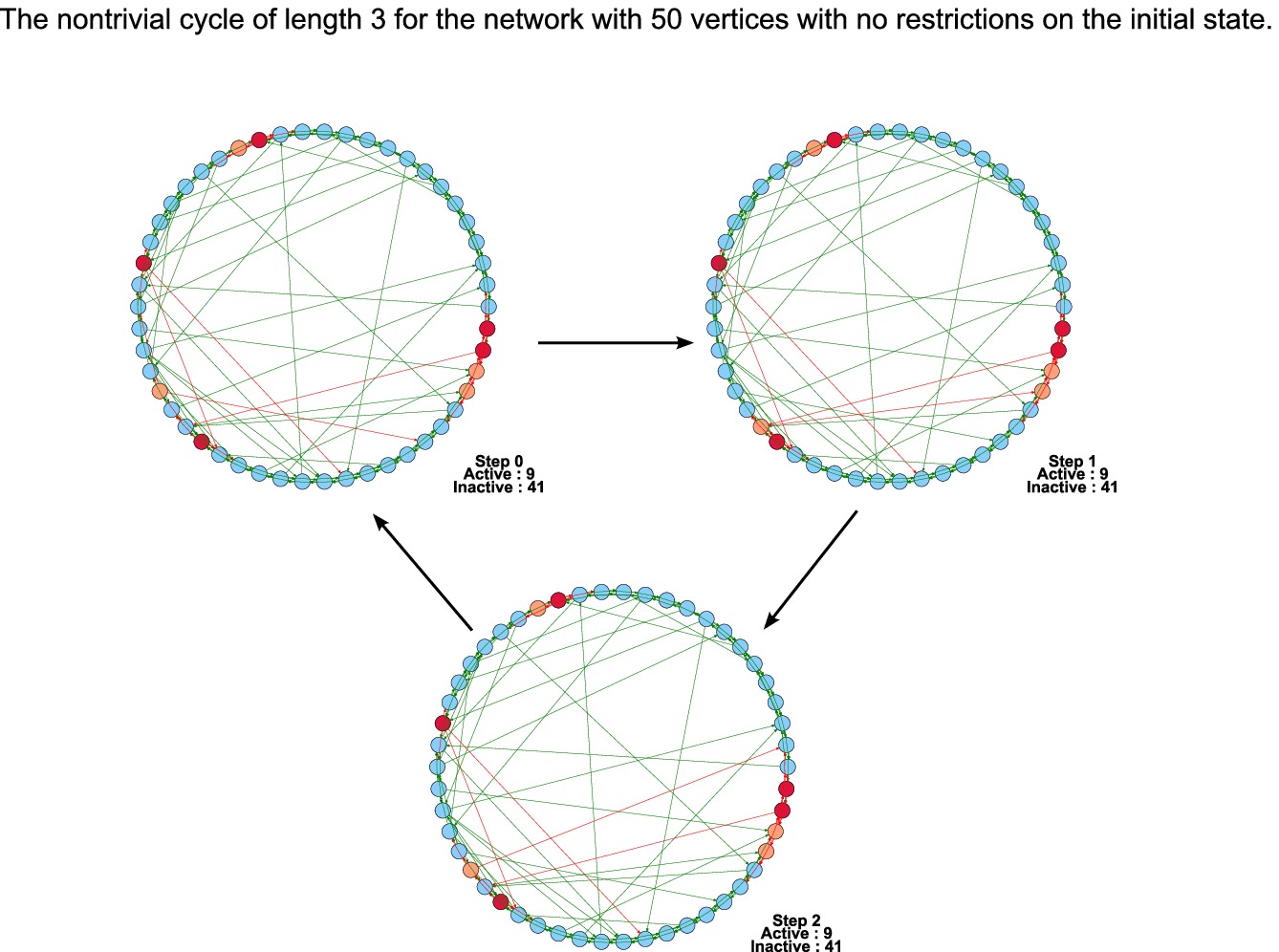}
\caption{
{\bf The nontrivial cycle of length 3 for the network of conformists with instigators.} 
At the initial state in the network there are both active and inactive simple agents.
}
\label{cycle2}
\end{figure}

Also it should be noted that the key condition in theorems 1 and 3 is that all simple agents must be either all inactive or all active at the initial time moment. If we drop this condition, the corresponding networks can display the behavior different from that described by Theorems 1 and 3. For example in Figure \ref{cycle2} we demonstrate the cycle of length $3$ for the network with instigators, where all simple agents are conformists, but at the initial state there are both active and inactive simple agents.

We would like to note that for the models proposed it is possible to study more complex dynamical properties using the formalism of quantified Boolean formulas with two quantification levels (2QBF) \cite{GMN09HBSAT}. Suppose that $\Phi$ is a disposition of instigators and $\Psi$ is a disposition of loyalists. Then, for example, the condition that there exists such disposition of instigators, that for any disposition of loyalists the network, starting from the state with inactive simple agents, after several time moments transitions to a state in which almost all simple agents are active, can be described using the 2QBF of the following kind:
\begin{equation*}
\exists \Phi \forall \Psi \Re\left(G,F_G, \Phi, \Psi\right)
\end{equation*}
This condition can be considered as an improved variant of condition describing the vulnerability of the network to instigators. To solve such problems one can use modern 2QBF-solvers \cite{GMN09HBSAT}, \cite{DBLP:conf/sat/JanotaS11}. We can also take into account any constraints on the cardinality of $\Phi$ and $\Psi$.

Finally, one natural extension of the proposed models is to assign various types of weights to network arcs and modify vertex weight functions accordingly. Arc weights can represent social pressure, authority, etc. for each particular member of a collective. In addition to that, it would be interesting to study the dynamics of networks in which weight function of a vertex can take into account the influence of vertices that are at a distance $>1$ in $G$ from the vertex considered. All the listed aspects can be quite easily implemented into corresponding SAT encodings. We plan to do it in the nearest future.

\section*{Acknowledgements}
We are thankful to Ilya Otpuschennikov for his help with constructing SAT encodings of the considered problems.

We also thank D.A. Novikov and V.V. Breyer for their comments and suggestions made during discussions on the early variants of the present research.

We express our deep gratitude to A.A. Evdokimov for attracting our attention to the study of dynamical properties of discrete automaton functions. It is the development of early ideas from \cite{EKS} that led us to the results of the present paper.


%
%
%

\end{document}